\documentclass[aps,prc,reprint,superscriptaddress]{revtex4-2}
\usepackage{amsmath,amssymb,amsfonts}
\usepackage{physics}
\usepackage{graphicx}
\begin{document}

\title{Tensor analyzing power $T_{20}$ in the reaction $\gamma \vec{d}\to pn$ at photon energies 350--680 MeV}

\author{\firstname{V.~V.}~\surname{Gauzshtein}}
\email{V.V.Gauzshteyn@inp.nsk.su}
\affiliation{%
Budker Institute of Nuclear Physics of Siberian Branch Russian Academy of Sciences, 630090 
Novosibirsk, Russia 
}%

\author{\firstname{A.~I.}~\surname{Fix}}
\email{A.I.Fiks@inp.nsk.su}
\affiliation{%
Budker Institute of Nuclear Physics of Siberian Branch Russian Academy of Sciences, 630090 
Novosibirsk, Russia 
}%
\affiliation{%
Novosibirsk State University, 630090 Novosibirsk, Russia 
}%

\author{\firstname{V.~I.}~\surname{Ivanov}}
\affiliation{%
Budker Institute of Nuclear Physics of Siberian Branch Russian Academy of Sciences, 630090 
Novosibirsk, Russia 
}%

\author{\firstname{D.~M.}~\surname{Nikolenko}}
\affiliation{%
Budker Institute of Nuclear Physics of Siberian Branch Russian Academy of Sciences, 630090 
Novosibirsk, Russia 
}%

\author{\firstname{I.~A.}~\surname{Rachek}}
\affiliation{%
Budker Institute of Nuclear Physics of Siberian Branch Russian Academy of Sciences, 630090 
Novosibirsk, Russia 
}%

\author{\firstname{Yu.~V.}~\surname{Shestakov}}
\affiliation{%
Budker Institute of Nuclear Physics of Siberian Branch Russian Academy of Sciences, 630090 
Novosibirsk, Russia 
}%
\affiliation{%
Novosibirsk State University, 630090 Novosibirsk, Russia 
}%

\author{\firstname{D.~K.}~\surname{Toporkov}}
\affiliation{%
Budker Institute of Nuclear Physics of Siberian Branch Russian Academy of Sciences, 630090 
Novosibirsk, Russia 
}%

\author{\firstname{A.~V.}~\surname{Yurchenko}}
\affiliation{%
Budker Institute of Nuclear Physics of Siberian Branch Russian Academy of Sciences, 630090 
Novosibirsk, Russia 
}%

\author{\firstname{S.~A.}~\surname{Zevakov}}
\affiliation{%
Budker Institute of Nuclear Physics of Siberian Branch Russian Academy of Sciences, 630090 
Novosibirsk, Russia 
}%

\author{\firstname{G.~N.}~\surname{Baranov}}
\affiliation{%
Budker Institute of Nuclear Physics of Siberian Branch Russian Academy of Sciences, 630090 
Novosibirsk, Russia 
}%

\author{\firstname{A.~V.}~\surname{Bogomyagkov}}
\affiliation{%
Budker Institute of Nuclear Physics of Siberian Branch Russian Academy of Sciences, 630090 
Novosibirsk, Russia 
}%

\author{\firstname{V.~M.}~\surname{Borin}}
\affiliation{%
Budker Institute of Nuclear Physics of Siberian Branch Russian Academy of Sciences, 630090 
Novosibirsk, Russia 
}%

\author{\firstname{E.~M.}~\surname{Darwish}}
\affiliation{%
Physics Department, College of Science, Taibah University, Medina, 41411, Saudi Arabia
}%
\affiliation{%
Physics Department, Faculty of Science, Sohag University, Sohag, 82524, Egypt
}%

\author{\firstname{D.~V.}~\surname{Dorokhova}}
\affiliation{%
Budker Institute of Nuclear Physics of Siberian Branch Russian Academy of Sciences, 630090 
Novosibirsk, Russia 
}%

\author{\firstname{V.~L.}~\surname{Dorokhov}}
\affiliation{%
Budker Institute of Nuclear Physics of Siberian Branch Russian Academy of Sciences, 630090 
Novosibirsk, Russia 
}%

\author{\firstname{S.~E.}~\surname{Karnaev}}
\affiliation{%
Budker Institute of Nuclear Physics of Siberian Branch Russian Academy of Sciences, 630090 
Novosibirsk, Russia 
}%

\author{\firstname{A.~A.}~\surname{Kovalenko}}
\affiliation{%
Budker Institute of Nuclear Physics of Siberian Branch Russian Academy of Sciences, 630090 
Novosibirsk, Russia 
}%

\author{\firstname{V.~N.}~\surname{Kudryavtsev}}
\affiliation{%
Budker Institute of Nuclear Physics of Siberian Branch Russian Academy of Sciences, 630090 
Novosibirsk, Russia 
}%

\author{\firstname{M.~I.}~\surname{Levchuk}}
\affiliation{%
Stepanov Institute of Physics, National Academy of Sciences of Belarus, Minsk, 220072, Belarus
}%
\affiliation{%
Institute of Applied Physics, National Academy of Sciences of Belarus, Minsk, 220072, Belarus
}%

\author{\firstname{E.~B.}~\surname{Levichev}}
\affiliation{%
Budker Institute of Nuclear Physics of Siberian Branch Russian Academy of Sciences, 630090 
Novosibirsk, Russia 
}%

\author{\firstname{I.~B.}~\surname{Logashenko}}
\affiliation{%
Budker Institute of Nuclear Physics of Siberian Branch Russian Academy of Sciences, 630090 
Novosibirsk, Russia 
}%
\affiliation{%
Novosibirsk State University, 630090 Novosibirsk, Russia 
}%

\author{\firstname{A.~Yu.}~\surname{Loginov}}
\affiliation{%
National Research Tomsk Polytechnical University, Tomsk, 634050,  Russia
}%
\affiliation{%
Tomsk State University of Control Systems and Radioelectronics, Tomsk, 634050, Russia
}%

\author{\firstname{T.~V.}~\surname{Maltsev}}
\affiliation{%
Budker Institute of Nuclear Physics of Siberian Branch Russian Academy of Sciences, 630090 
Novosibirsk, Russia 
}%

\author{\firstname{R.~Z.}~\surname{Mamutov}}
\affiliation{%
Budker Institute of Nuclear Physics of Siberian Branch Russian Academy of Sciences, 630090 
Novosibirsk, Russia 
}%

\author{\firstname{I.~N.}~\surname{Okunev}}
\affiliation{%
Budker Institute of Nuclear Physics of Siberian Branch Russian Academy of Sciences, 630090 
Novosibirsk, Russia 
}%

\author{\firstname{P.~A.}~\surname{Piminov}}
\affiliation{%
Budker Institute of Nuclear Physics of Siberian Branch Russian Academy of Sciences, 630090 
Novosibirsk, Russia 
}%

\author{\firstname{L.~I.}~\surname{Shekhtman}}
\affiliation{%
Budker Institute of Nuclear Physics of Siberian Branch Russian Academy of Sciences, 630090 
Novosibirsk, Russia 
}%

\author{\firstname{E.~A.}~\surname{Simonov}}
\affiliation{%
Budker Institute of Nuclear Physics of Siberian Branch Russian Academy of Sciences, 630090 
Novosibirsk, Russia 
}%

\author{\firstname{S.~V.}~\surname{Sinyatkin}}
\affiliation{%
Budker Institute of Nuclear Physics of Siberian Branch Russian Academy of Sciences, 630090 
Novosibirsk, Russia 
}%

\author{\firstname{M.~A.}~\surname{Skamarokha}}
\affiliation{%
Budker Institute of Nuclear Physics of Siberian Branch Russian Academy of Sciences, 630090 
Novosibirsk, Russia 
}%

\author{\firstname{E.~V.}~\surname{Starostina}}
\affiliation{%
Budker Institute of Nuclear Physics of Siberian Branch Russian Academy of Sciences, 630090 
Novosibirsk, Russia 
}%

\author{\firstname{I.~A.}~\surname{Ulev}}
\affiliation{%
Budker Institute of Nuclear Physics of Siberian Branch Russian Academy of Sciences, 630090 
Novosibirsk, Russia 
}%

\author{\firstname{A.~N.}~\surname{Zhuravlev}}
\affiliation{%
Budker Institute of Nuclear Physics of Siberian Branch Russian Academy of Sciences, 630090 
Novosibirsk, Russia 
}%

\date{\today}

\begin{abstract}
We present measurements of the tensor analyzing power $T_{20}$ in the reaction $\gamma {d}\to pn$ at photon energies $E_\gamma = 350$--$680$~MeV. The experiment was performed at the VEPP-3 storage ring using an internal tensor-polarized deuterium gas target and a tagged quasi-real photon beam. The data were obtained for proton emission angles $\Theta_p = 70^\circ$--$102^\circ$. Comparison with modern meson-baryon calculations shows generally satisfactory agreement within the present uncertainties. A theoretical analysis of an extended data set, covering the energy range from threshold to 680~MeV and including both the new data and the previous results from 2007, is presented.
\end{abstract}

\maketitle

\section{Introduction}
\label{sec:intro}
The noncentral (tensor) components of the nuclear interaction are known to play a crucial role in electro- and photonuclear reactions, particularly at short internucleon distances~\cite{nnreview}. However, detailed experimental investigations of these effects have emerged only in recent years. For a long time, the main obstacle was the lack of tensor-polarized targets with a high degree of polarization suitable for use with storage rings or extracted beams. Without such targets, isolating tensor effects, measuring analyzing powers, and separating spin amplitudes were practically impossible.

The shift from unpolarized cross sections to spin observables in the 1980s opened the way for direct tests of the deuteron structure, particularly its connection to the tensor component of the $NN$ forces. One of the key technological breakthroughs was the development of internal polarized gas targets for charged-particle storage rings, first implemented at the Budker Institute of Nuclear Physics (Novosibirsk)~\cite{dimitriev85,wojtse86}. The implementation of atomic beam sources and, subsequently, storage cells at the VEPP-2, VEPP-3, NIKHEF, and Bates facilities~\cite{gilman90,bouwhuis99,the91} allowed the effective thickness of the deuterium target to be significantly increased while maintaining a high degree of tensor polarization ($P_{zz} \approx 0.6$--$0.8$). Furthermore, it enabled rapid switching of the polarization sign, which is crucial for minimizing systematic uncertainties caused by the drift in detector and beam characteristics. Parallel developments in recoil polarization measurement techniques (such as the POLDER and AHEAD detectors)~\cite{abbott00,garcon94} and coincidence methods radically improved the signal-to-background ratio. These advancements also ensured precise photon energy control at the level of $\Delta E_\gamma / E_\gamma < 10^{-3}$, which is critically important given the rapid variation of observables with increasing momentum transfer.

An alternative approach is pursued at the Thomas Jefferson National Accelerator Facility (JLab), where an extensive experimental program focuses on measuring tensor structure functions (such as $b_1$ and TMDs) using solid deuterated ammonia ($\mathrm{ND}_3$) targets~\cite{Slifer2013,Poudel2025}. Polarized via dynamic nuclear polarization, these targets can now achieve tensor polarization levels of $\sim 30\%$ under beam conditions. This level of polarization provides the high luminosity required for precision measurements at momentum transfers up to $Q^2 \sim 5$~GeV$^2$, where gas targets become statistically inefficient. However, such solid targets inherently contain a large fraction of unpolarized nuclei (nitrogen, and the surrounding helium bath), leading to a significant dilution factor that requires careful background subtraction. In this regard, current experiments with internal gas targets at VEPP-3, utilizing pure polarized deuterium free from such dilution and allowing for rapid switching of the polarization sign ($\sim 30$~s), provide superior control over systematic uncertainties.

Among the three independent components of the tensor analyzing power $T_{2M}$ ($M=0,1,2$), $T_{20}$ is the most experimentally accessible for the reaction $\gamma d \to pn$. Reliable measurement of the other two components, $T_{21}$ and $T_{22}$, generally requires the use of a transversely polarized target (which is usually technically challenging) or an extremely precise reconstruction of the azimuthal angles of the emitted nucleons, which inevitably leads to an increase in systematic uncertainties.

Turning to the theoretical description of deuteron photodisintegration, it is well known from classic works (see, e.g., the review \cite{ArenhFBS}) that the single-nucleon mechanism in the $(\gamma, p)$ reaction is strongly suppressed by conservation laws, since a real photon cannot be absorbed by a free nucleon without violating energy and momentum balance. Consequently, the process is forced to occur on off-shell nucleons carrying high internal momenta ($p \gtrsim 300$~MeV/$c$ already at $E_\gamma \sim 100$~MeV). In such kinematics, two-nucleon mechanisms, primarily driven by the exchange of virtual mesons, play a dominant role. This naturally identifies meson-baryon degrees of freedom as the relevant ones, and the corresponding description is believed to remain adequate at least up to the region around 1~GeV. 

Below the pion production threshold, the meson-baryon approach provides a reasonably good description of the data, provided a realistic $NN$-potential is used and gauge invariance of the full electromagnetic current is maintained. At energies above the pion threshold ($E_\gamma > 140$~MeV), the opening of the inelastic $\pi NN$ channel transforms the $\Delta$-isobar from a virtual configuration into a real dynamic degree of freedom, requiring a three-body $\pi NN$ treatment. This generally leads to a significant complication of the theoretical description. Overall, compared to low-energy calculations, theoretical approaches in the region $E_\gamma > 140$~MeV remain less complete and are more phenomenological in nature. This points to the need for new experimental data to better constrain existing models and identify the dominant reaction mechanisms.

Regarding nucleon resonances lying above $\Delta(1232)\frac{3}{2}^+$, it is evident that their influence should increase with energy. Calculations \cite{SWA}, including $N(1520)\frac{3}{2}^-$ and $N(1535)\frac{1}{2}^-$, show that their contribution to the total cross section remains moderate even in the second resonance region (about 14\% at $E_\gamma=680$~MeV). However, polarization observables, being sensitive to the interference between different spin structures, may exhibit a stronger response to their presence. Therefore, expanding polarization measurements in the second and third resonance regions would help clarify the role of nucleon excitations in deuteron photodisintegration.

While the meson-baryon framework is expected to provide an adequate description of deuteron photodisintegration, there has been active discussion regarding the limits of its applicability~\cite{BrodskyHiller1992,Farrar1994}.  Specifically, it remains unclear at which energies quark degrees of freedom begin to play a role, thereby rendering more traditional approaches inadequate. On general grounds, one expects this transition to occur well above the GeV scale. Indeed, given $\Lambda_{\text{QCD}} \approx 250$~MeV, the onset of the perturbative QCD regime, characterized by the fulfillment of the constituent counting rules and helicity conservation, requires hard momentum transfers $|t| \gtrsim 2$--$5$~GeV$^2$ and photon energies $E_\gamma \gtrsim 3$--$4$~GeV~\cite{schulte01}, a range that clearly exceeds the photon energies relevant to the present study ($E_\gamma \le 700$~MeV). 

However, the experimental situation at intermediate energies is far more complex. Particularly instructive in this regard are the recoil-polarization measurements in the $\gamma d \to pn$ reaction~\cite{wije01}, where the induced polarization $P_y$ was found to deviate systematically from the predictions of purely meson-baryon models~\cite{kang}, the values measured above $E_\gamma \sim 1$~GeV being small and consistent with zero. Taken alone, such behavior could be regarded as an approach to the asymptotic pQCD limit $P_y = 0$, yet the polarization-transfer components $C_x$ and $C_z$ measured in the same experiment contradict the helicity-conservation rule, indicating that the smallness of $P_y$ has an origin unrelated to perturbative dynamics. These observations lend support to the view that the energy region around 1~GeV represents a transition regime in which neither purely hadronic nor pQCD descriptions are fully adequate. In this context, various hybrid models combining traditional nuclear and quark-gluon degrees of freedom are being actively explored. This further underscores the need to develop theoretical models and conduct new precision experiments for their reliable verification. 

To expand the experimental database in the intermediate energy region ($E_\gamma \sim 300$--$1200$~MeV), precision measurements are being carried out at the Budker Institute of Nuclear Physics. This program utilizes an internal tensor-polarized deuterium gas target and a tagged photon beam. To date, the only measurements of $T_{2M}$ in the reaction $\gamma d \to np$ were performed at the VEPP-3 facility in 2007~\cite{Rachek2007}. Subsequently, the VEPP-3 detector system was significantly upgraded. In particular, a photon tagging system and a wide-aperture sandwich detector for neutron identification were implemented~\cite{Shest2024}. This has led to much cleaner event selection, particularly at $E_\gamma > 500$~MeV, where the background increases rapidly. In this paper, we report the first $T_{20}$ measurements in $\gamma\vec{d}\to pn$ at photon energies $350$--$680$~MeV using the upgraded detector and tagging system.

The paper is organized as follows. Section~\ref{sec:setup} describes the DEUTERON experimental setup at the VEPP-3 storage ring, the method for producing a quasi-real tagged photon beam, and the internal tensor-polarized gas target. Section~\ref{sec:data} briefly outlines the procedure for extracting the $T_{20}$ component from the measured polarization asymmetry. Section~\ref{sec:model} presents the theoretical model used to analyze the data. Section~\ref{sec:results} reports the measurement results and their comparison with modern meson-baryon models. Finally, Sec.~\ref{sec:conclusion} summarizes the main findings and outlines prospects for further development of the experimental program.

\section{Experimental setup}
\label{sec:setup}

The measurement of the tensor analyzing power $T_{2M}$ ($M=0,1,2$) in the reaction $\gamma \vec{d} \to pn$ was performed at the DEUTERON setup located in the straight section of the VEPP-3 storage ring with an electron beam energy of $E_e = 800$~MeV. The internal target consists of a thin-walled storage cell continuously fed with polarized deuterium gas from an atomic beam source (ABS). Injecting polarized atoms into the cooled internal cell increases the effective target thickness by a factor of $\sim 100$ compared to a jet target, reaching $\sim 5 \times 10^{13}$~atoms/cm$^2$. The sign of the tensor polarization is automatically reversed every 30~s, enabling the accumulation of statistics for two polarization states and the measurement of the polarization asymmetry. 

Although the ABS output polarization is close to 100\%, wall collisions, atom-atom, and electron-beam interactions reduce it inside the cell. The average tensor polarization $P_{zz}$ was monitored using a ``Low-$Q$'' polarimeter, which operates by measuring the asymmetry of elastic $ed$ scattering at low momentum transfers, utilizing precisely calculated theoretical values of $T_{20}$. The resulting values are $P_{zz} = 0.39 \pm 0.03$ for the configuration with the ideal ABS polarization $P_{zz}^{\mathrm{ABS}} = +1$, and $P_{zz} = -0.66 \pm 0.05$ for $P_{zz}^{\mathrm{ABS}} = -2$. A more detailed description of the target and the polarization system can be found in Ref.~\cite{Shest2024}.

\begin{figure}[t!]
\centering
\includegraphics[width=1.0\columnwidth]{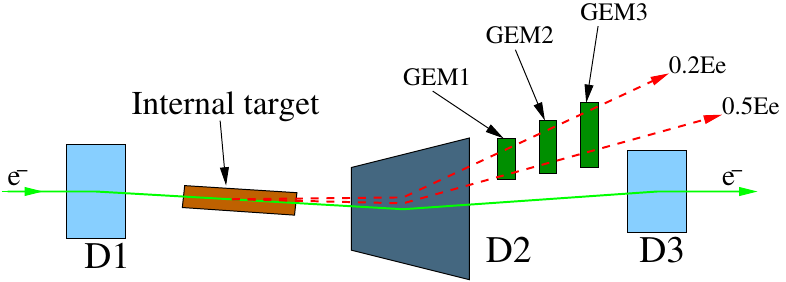}
\caption{
Schematic diagram of the photon tagging system (PTS). The electron beam with an initial energy $E_e$ (in this experiment, $E_e = 800$~MeV) is deflected by magnet D1 before the internal target; scattered electrons are extracted from the orbit by magnet D2 and detected by GEM detectors (GEM1--GEM3), while unscattered electrons are returned to the original orbit by magnet D3. The dashed lines indicate example trajectories of scattered electrons corresponding to specific tagged photon energies.}
\label{fig:PTS}
\end{figure}

The $\gamma d \to pn$ reaction was induced by quasi-real bremsstrahlung photons produced by electron scattering at angles less than $1^\circ$ ($Q^2 \sim 0$). The photon energy was reconstructed using a photon tagging system (PTS) based on dipole magnets and fast GEM trackers (Fig.~\ref{fig:PTS}), achieving an energy resolution of $\sim 1\%$. For the electron beam energy of $E_e = 800$~MeV used in the present experiment, the tagged photon energy ranged from $0.44\,E_e$ to $0.85\,E_e$ ($E_\gamma = 350$--$680$~MeV).

\begin{figure}[t!]
\centering
\includegraphics[width=1.0\columnwidth]{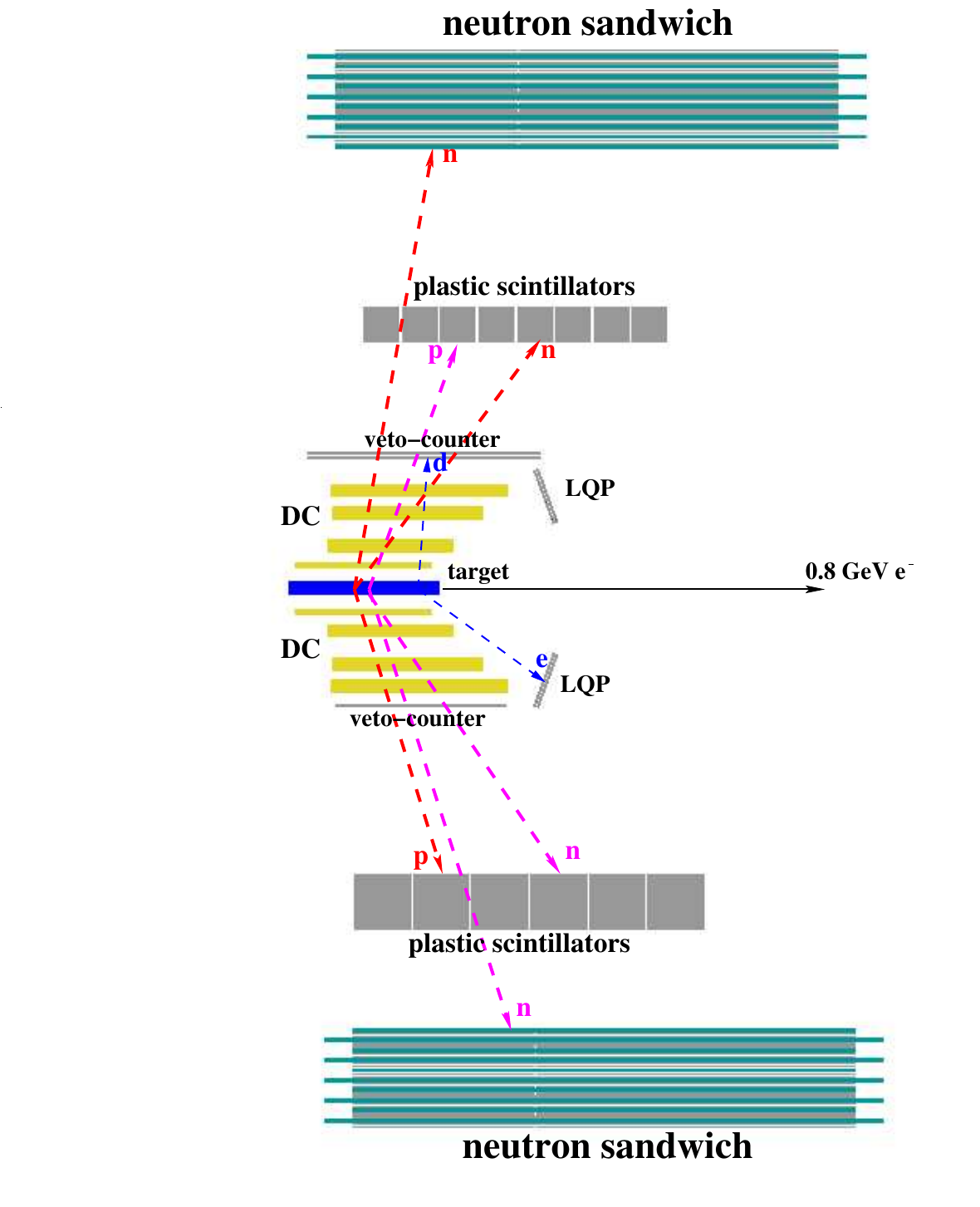}
\caption{
Schematic view of the detector system used in the experiment. The system comprises two symmetric arms arranged above and below the target. Each arm contains drift chambers (DC) for charged particle track reconstruction, veto counters to distinguish between charged and neutral particles, plastic scintillators for proton and neutron detection, and neutron sandwich detectors consisting of alternating layers of iron and scintillator for neutron detection. The tensor polarization of the target is monitored using a Low-$Q$ polarimeter (LQP), which consists of two symmetrically arranged sets of scintillation counters for detecting electrons and deuterons from elastic $ed$ scattering. Characteristic trajectories of protons (p), neutrons (n), electrons (e), and deuterons (d) are shown. The $0.8$-GeV electron beam travels from left to right.}
\label{fig:setup}
\end{figure}

The detector system consists of two identical arms arranged symmetrically above and below the target (Fig.~\ref{fig:setup}). Each arm includes drift chambers for charged-particle track reconstruction, thin scintillation veto counters for background suppression and charge identification, as well as scintillator hodoscopes and a segmented hadronic sandwich (iron/scintillator) array for neutron detection. The experimental setup required the coincidence detection of a tagged photon, a proton, and a neutron in a cross-arm configuration (proton in the upper arm and neutron in the lower arm, or vice versa). Particle identification employed time-of-flight (TOF) and energy-loss ($\Delta E/E$) measurements using two-layer plastic scintillators for protons, while for neutrons it relied on veto signals and calorimetric measurements.

\section{Event Selection}
\label{sec:data}

When the polarization axis (aligned with the external magnetic field) coincides with the direction of the quasi-real photon beam, the differential cross section can be expressed in terms of the tensor analyzing power component $T_{20}$ as
\begin{equation}\label{eq:dsigma_simple}
\frac{d\sigma}{d\Omega} = \frac{d\sigma_0}{d\Omega} \left[ 1 + \frac{1}{\sqrt{2}} P_{zz} T_{20} \right],
\end{equation}
where $d\sigma_0/d\Omega$ is the cross section for an unpolarized target. The number of detected events $N$ is proportional to the cross section:
\begin{equation}\label{eq:N_events}
N^{\pm} = \mathcal{L} \cdot \frac{d\sigma_0}{d\Omega} \left[ 1 + \frac{1}{\sqrt{2}} P_{zz}^{\pm} T_{20} \right],
\end{equation}
where $\mathcal{L}$ is the integrated luminosity (identical for both polarization states due to the rapid switching), and $P_{zz}^{+}$ and $P_{zz}^{-}$ are the values of the target tensor polarization in the two different states. The value of $T_{20}$ was extracted from the experimental yields $N^{\pm}$, accounting for the measured values of $P_{zz}^{\pm}$, using the formula that follows directly from Eq.~(\ref{eq:N_events}):
\begin{equation}
T_{20} = \sqrt{2} \, \frac{N^{+} - N^{-}}{P_{zz}^{+} N^{-} - P_{zz}^{-} N^{+}}.
\label{eq:T20_exact}
\end{equation}

\begin{figure}[t!]
\centering
\includegraphics[width=1.0\columnwidth]{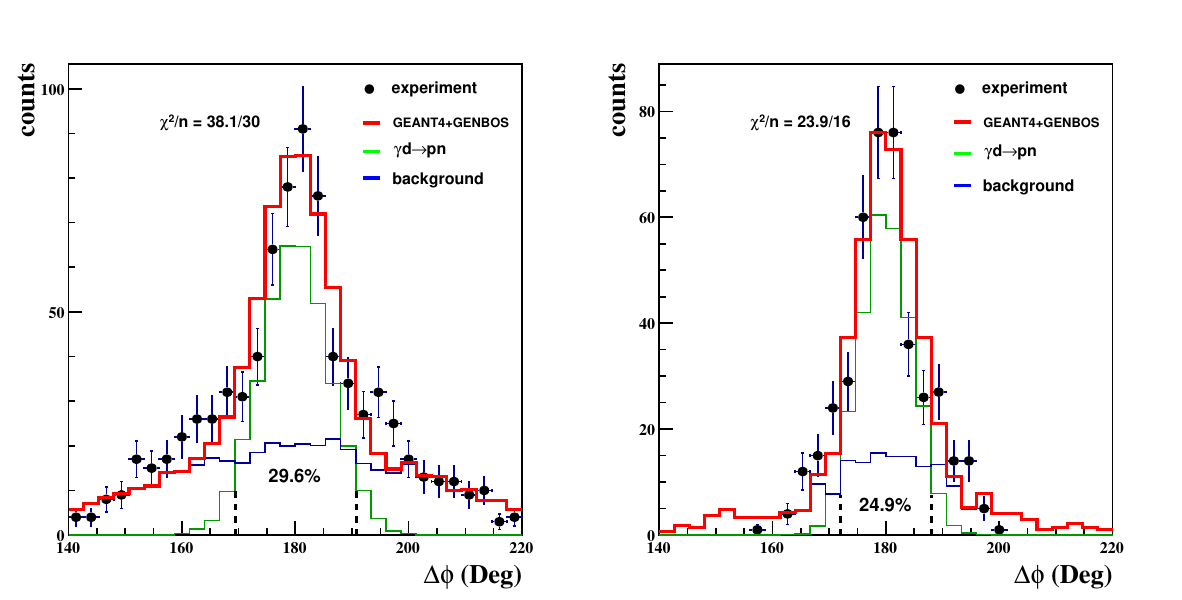}
\caption{
Distribution of the azimuthal angle difference $\Delta\Phi=\Phi_p-\Phi_n$ for events where both the proton and the neutron are detected by scintillation spectrometers. Left panel: the proton is detected in the lower arm and the neutron in the upper arm. Right panel: the proton is detected in the upper arm and the neutron in the lower arm. Points: experimental data. Curves: results of the GEANT4 simulation; the green curve corresponds to the $\gamma d \to pn$ reaction, the blue curve shows the contribution from background reactions ($\gamma d \to \pi^-\pi^+ pn$, $\gamma d \to \pi^0 pn$, and $\gamma d \to \pi^0 \pi^0 pn$), and the red curve represents the sum of all reactions.
}
\label{fig:dPhi_scint}
\end{figure}

\begin{figure}[t!]
\centering
\includegraphics[width=1.0\columnwidth]{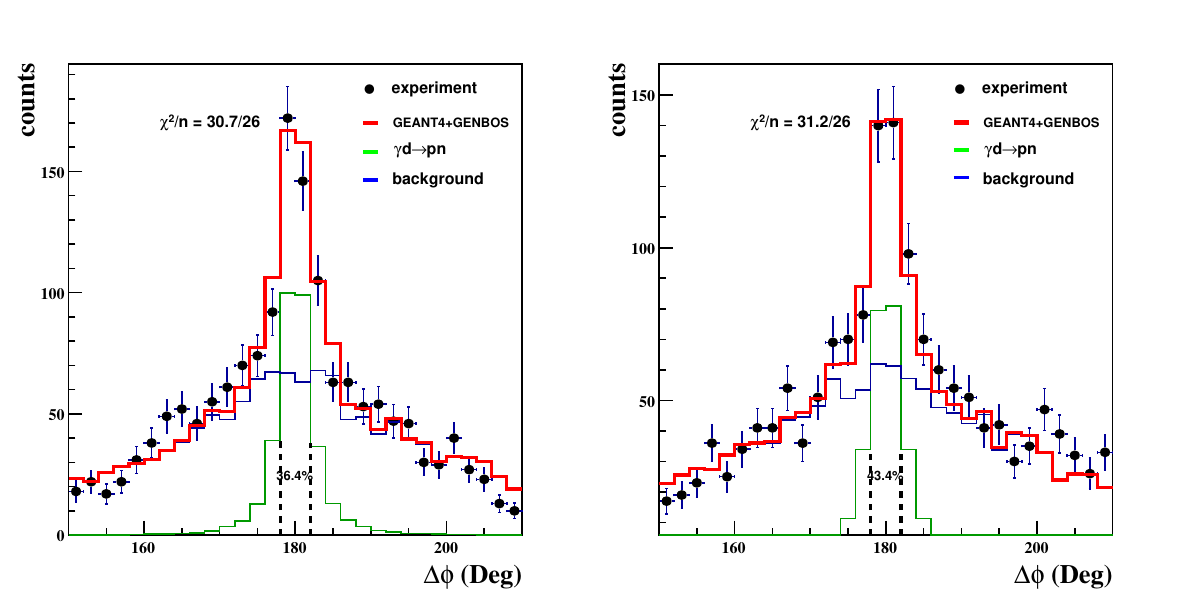}
\caption{
Same as in Fig.~\ref{fig:dPhi_scint}, but for the case where the neutron was detected by the sampling iron/scintillator detector.}
\label{fig:dPhi_sandwich}
\end{figure}

The measurement of $T_{20}$ according to Eq.~(\ref{eq:T20_exact}) was carried out in full coincidence kinematics. The primary criterion for identifying true deuteron photodisintegration events ($\gamma d \to pn$) was the correlation between the azimuthal emission angles of the particles: $\Delta\Phi = |\Phi_p - \Phi_n| \approx 180^\circ$. The accumulated experimental $pn$-coincidence sample was divided into two groups depending on the types of detectors involved. In the first case, both the proton and the neutron were detected by scintillation spectrometers (in either the ``upper--lower'' or ``lower--upper'' configurations). In the second case, the proton was detected by a scintillation spectrometer, while the neutron was detected by the sampling iron/scintillator detector. The $\Delta\Phi$ distributions for the selected events in both groups are shown in Figs.~\ref{fig:dPhi_scint} and~\ref{fig:dPhi_sandwich}.


The polar ($\Theta_p$) and azimuthal ($\Phi_p$) emission angles of the protons were measured using drift chambers with an angular resolution of better than $0.6^\circ$. The azimuthal angle of the neutrons detected by the scintillation spectrometers was determined from the difference in light arrival times at opposite ends of the scintillator. The angular resolution did not exceed $6^\circ$ for the upper spectrometer and $7^\circ$ for the lower one. For neutrons detected by the hadronic sandwich system, the azimuthal angle was reconstructed from the coordinates of the scintillator strips where energy deposition occurred. In the present analysis, only events with energy deposition in at most two strips were included.

With $\Delta\Phi = 180^\circ$ by kinematics for true photodisintegration events, the observed peak shape is governed primarily by the finite angular resolution in $\Phi_n$ and by an irreducible background dominated by the $\gamma d \to \pi^0 pn$ reaction. To validate the event selection criteria based on $\Delta\Phi$ and to quantitatively estimate the background, the experimental distributions were compared with Monte Carlo simulation results obtained using the GEANT4 package and the GENBOS photoreaction generator. The geometry of the setup and the simulation conditions exactly matched the experimental ones. The GENBOS generator describes all the main reaction channels required for this analysis, with the exception of the $\gamma d \to \pi^0\pi^0 d$ process, whose contribution to the selected sample is negligible. The satisfactory agreement between the experimental data and the simulation results ($\chi^2/\mathrm{ndf} < 2$) verifies our methodology for identifying $pn$ coincidences and reconstructing the azimuthal emission angles.

\section{Model}
\label{sec:model}

\begin{figure}[ht!]
\centering
\includegraphics[width=0.8\columnwidth]{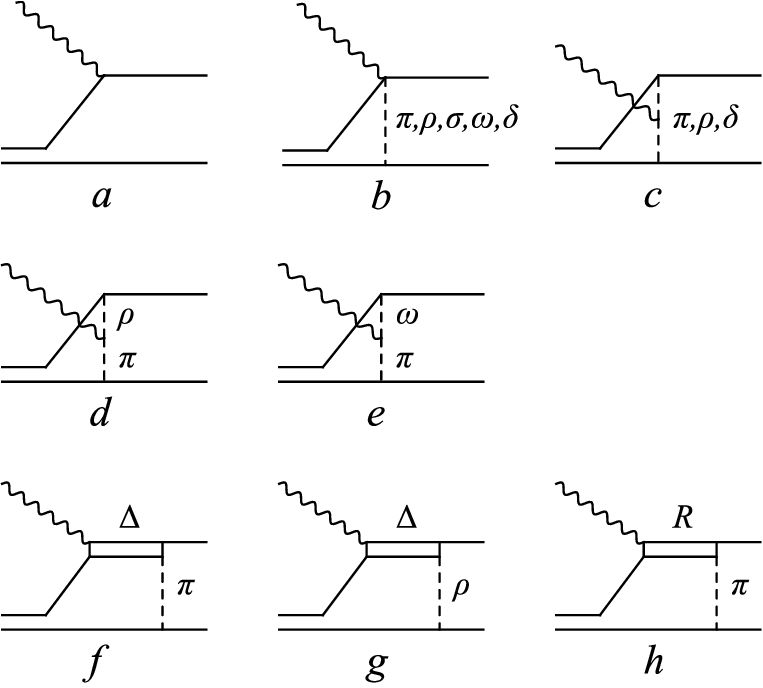}
\caption{
Diagrammatic representation of the mechanisms of the deuteron photodisintegration reaction $\gamma d \to np$ included in this calculation. The $\Delta(1232)\,\frac{3}{2}^+$ isobar is denoted by the symbol $\Delta$ (diagrams $f$ and $g$). The resonance states $N(1440)\,\frac{1}{2}^+$, $N(1520)\,\frac{3}{2}^-$, and $N(1535)\,\frac{1}{2}^-$ are denoted by the symbol $R$ (diagram $h$).}
\label{fig:diagrams}
\end{figure}

For the theoretical analysis of the experimental data, a relatively simple model for $\gamma d\to np$ was employed, in which the reaction amplitude was calculated in the momentum representation, accounting for the dominant interaction mechanisms \cite{Laget,Levchuk}. The calculations were carried out within a standard nonrelativistic approach, taking into account the nuclear one-body current, meson-exchange currents, and isobar excitations. The corresponding diagrams are presented in Fig.~\ref{fig:diagrams}. A slightly modified version of this model, incorporating a three-body treatment of the intermediate $\Delta N$ interaction, was used in Ref.~\cite{Fix}. The meson-exchange current contributions include the exchange of pseudoscalar pions, scalar $\sigma$ and $\delta$ mesons, and vector $\rho$ and $\omega$ mesons. The corresponding parameters (meson masses, hadronic coupling constants, cutoff momenta, etc.) are adopted from the Bonn one-boson-exchange model for the $NN$ interaction \cite{OBEP} (OBEPQ version). Note that the seagull terms for the isoscalar mesons ($\sigma$, $\omega$) are associated with the spin-orbit component of the nucleon-nucleon potential. They are of relativistic origin (of order $1/M^2$, where $M$ is the nucleon mass) and emerge from the nonrelativistic reduction of the Lagrangian when intermediate antinucleon states are retained \cite{Riska}. The resonance sector includes the $\Delta(1232)\frac{3}{2}^+$ and three nucleon resonances ($N(1440)\frac{1}{2}^+$, $N(1520)\frac{3}{2}^-$, $N(1535)\frac{1}{2}^-$) accessible in the present energy range $E_\gamma<700$~MeV. For the description of the deuteron wave function, we used a separable representation of the Paris $NN$ potential \cite{Lacombe}, specifically the PEST4 version implemented in Ref.~\cite{Haiden}.

To account for the final-state $np$ interaction, we employed the standard scheme:
\begin{equation}\label{eq:FSI}
T=T_B+T_B\tau T_{NN}\,,
\end{equation}
where the Born term $T_B$ is the amplitude depicted by the diagrams in Fig.~\ref{fig:diagrams}, $\tau$ is the propagator of two free nucleons, and $T_{NN}$ is the $NN$ scattering $t$-matrix. For $T_{NN}$, a partial-wave expansion $T^{j(ls)}_{NN}$ was used, including states up to $^{2s+1}l_j = \,^3D_3$. We note that although the $s$-wave components are known to dominate the $NN$ interaction, taking into account configurations with $l>0$ is critically important for describing polarization observables.

As in the case of the deuteron wave function, the separable representation of the Paris potential~\cite{Lacombe} from Ref.~\cite{Haiden} was used to calculate the partial-wave components $T^{j(ls)}_{NN}$. In the region $E_\gamma > 300$~MeV, where the corresponding $NN$ kinetic energy in the laboratory frame exceeds 500~MeV, the model~\cite{Lacombe} is, in the strict sense, outside its range of applicability. However, as our results show, the contribution of this region to the final results is insignificant (see the data analysis in Sec.~\ref{sec:results}).

\begin{figure}[t!]
\centering
\includegraphics[width=1.0\columnwidth]{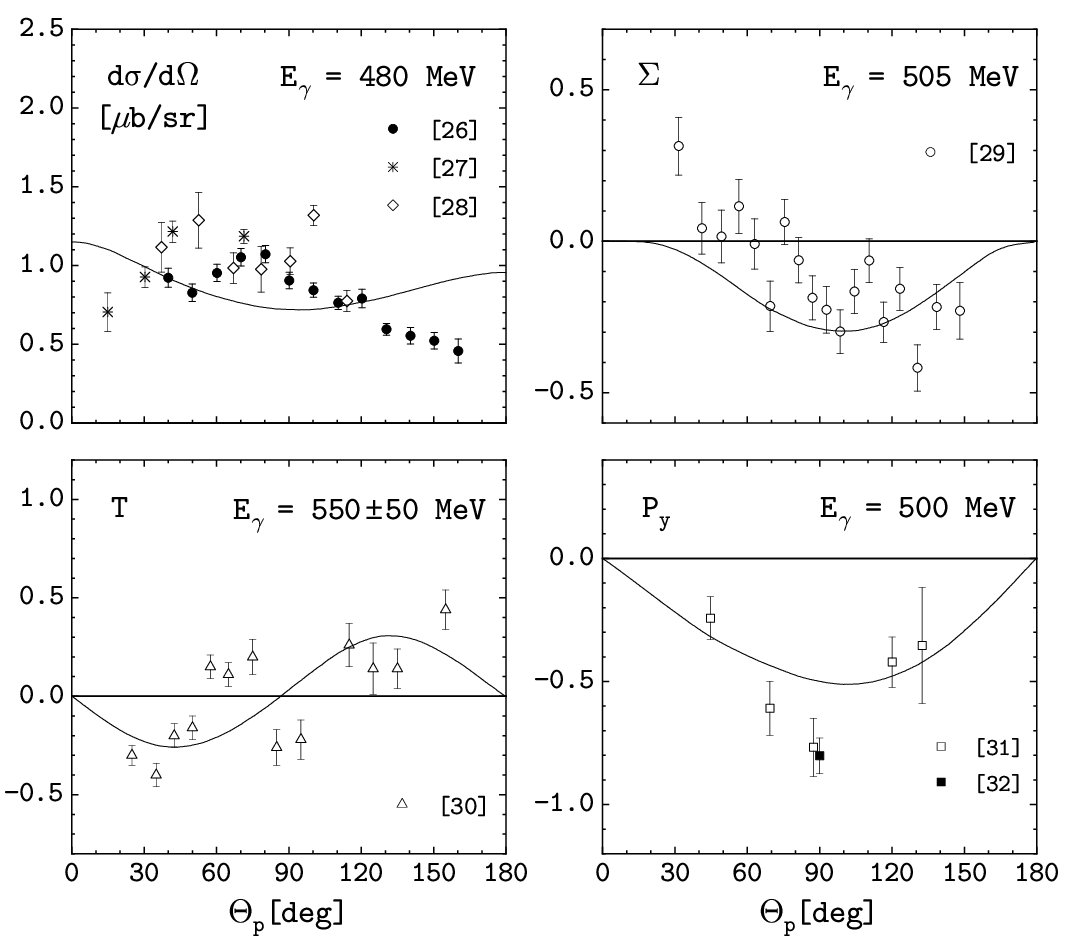}
\caption{
Comparison of our model calculations (solid curves) with experimental data for various observables in the $\gamma d \to pn$ reaction in the energy region above the $\Delta$-resonance. Top left: differential cross section $d\sigma/d\Omega$ at $E_\gamma = 480$~MeV. Top right: photon asymmetry $\Sigma$ at $E_\gamma = 505$~MeV. Bottom left: target asymmetry $T$ at $E_\gamma = 550\pm50$~MeV. Bottom right: recoil proton polarization $P_y$ at $E_\gamma = 500$~MeV. Experimental data are taken from the references indicated in the legends~\cite{Crawford:1996ka,Baba:1983yt,Dougan:1975nz,Bashkanov:2018ftd,Althoff:1984jjc,Ikeda:1978ika,Kamae:1976as}. 
}
\label{figObs}
\end{figure}

Our model closely follows the approach of Levchuk~\cite{Levchuk} by adding only three extra resonances (diagram $h$). When restricted to diagrams $(a\text{--}g)$, our results are in good agreement with those of Ref.~\cite{Levchuk}. The minor discrepancies are likely due to the use of a different parameterization of the $NN$ potential in Ref.~\cite{Levchuk} (the OBEPR version of the Bonn potential). 
As an illustration, Figure~\ref{figObs} demonstrates the agreement between our model and the experimental data for selected observables in the energy region $E_\gamma \sim 500$~MeV. As can be seen, the model generally describes the data quite well, although the characteristic shape of the differential cross section with a maximum around $\Theta_p \approx 80^\circ$ is not reproduced. This shortcoming, however, seems to be a common feature of most models in the considered energy region (see, e.g., \cite{ArenhFBS}).

Among the various mechanisms, the dominant contribution in the energy region of interest is expected to arise from meson exchange currents (diagrams $(b)$, $(c)$, $(d)$, and $(e)$ in Fig.~\ref{fig:diagrams}), with the pion current naturally playing the leading role. These terms are directly connected to the Kroll-Ruderman mechanism and the pion pole, leading primarily to electric dipole ($E1$) transitions. As for the single-nucleon term (diagram $(a)$), it is known to fall off rapidly with increasing energy~\cite{ArenhFBS}, a behavior dictated by the large spatial extent of the deuteron. In contrast, the meson-exchange currents are primarily sensitive to the radius of the $NN$ interaction, thereby falling off much more slowly and remaining significant at higher energies.

\begin{figure}[h!]
\centering
\includegraphics[width=1.0\columnwidth]{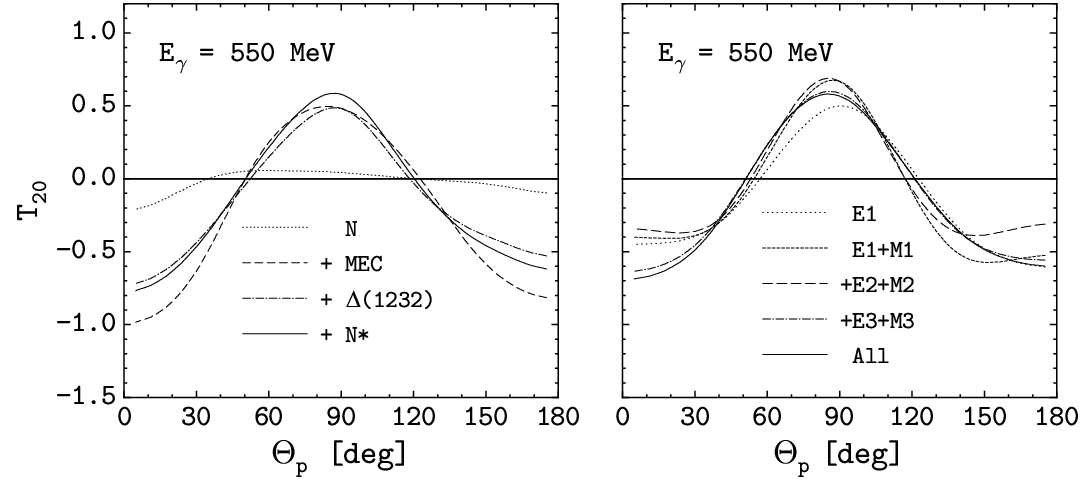}
\caption{
Tensor analyzing power $T_{20}$ for $\gamma d \to np$ at $E_\gamma = 550$~MeV. Left panel: contributions of individual mechanisms. The curves correspond to the one-nucleon current (diagram $(a)$ in Fig.~\ref{fig:diagrams}, dotted) with the successive addition of meson exchange currents (diagrams $(b)\text{--}(e)$, dashed), $\Delta(1232)$ excitation (diagrams $(f)$ and $(g)$, dash-dotted), and nucleon resonances $N(1440)\,\frac{1}{2}^+$, $N(1520)\,\frac{3}{2}^-$, and $N(1535)\,\frac{1}{2}^-$ (diagram $(h)$, solid). Right panel: convergence of the multipole expansion. The curves show the successive inclusion of electric and magnetic multipoles: only $E1$ (dotted), $E1+M1$ (short-dashed), $+E2+M2$ (long-dashed), $+E3+M3$ (dash-dotted), and the full result without multipole expansion (solid).
}
\label{fig:T20_Mech}
\end{figure}

These considerations are corroborated by direct calculations, shown in the left panel of Fig.~\ref{fig:T20_Mech}, which illustrates the role of various mechanisms in determining both the magnitude and the angular dependence of $T_{20}(\Theta_p)$. It is evident that the single-nucleon current (dotted curve) fails to describe the behavior of this observable even qualitatively. On the other hand, the inclusion of only meson (primarily pion) exchange currents (dashed curve), even without isobar configurations, provides a quite satisfactory description of both the magnitude and the shape of the asymmetry at intermediate angles, with noticeable discrepancies appearing only at the most forward and backward angles.

The right panel of the same figure demonstrates the role of various electric ($EL$) and magnetic ($ML$) multipoles in the formation of $T_{20}$. One can observe that in our angular region $\Theta_p = [70^\circ, 102^\circ]$, the $E1$ transition is dominant. Contributions from higher multipoles primarily serve as minor corrections. This result follows from the general properties of dipole radiation and is, in this sense, entirely expected. Indeed, in the classical picture, the incident electric field of the photon $\vec{E}$ induces a dipole moment $\vec{P}$ in the deuteron. The emitted proton then exhibits the angular dependence characteristic of dipole radiation, yielding $d\sigma/d\Omega \propto \sin^2\Theta_p$. Consequently, the $E1$ transition inevitably dominates near $\Theta_p \approx 90^\circ$ (unless it is suppressed by specific dynamical reasons), which is precisely confirmed by our calculations.

Thus, the primary role in deuteron photodisintegration at energies above 400~MeV is played by the same mechanisms that dominate the unpolarized cross section in the first resonance region, namely the pion exchange current and the $\Delta(1232)\frac{3}{2}^+$ excitation. The $N(1440)\frac{1}{2}^+$, $N(1520)\frac{3}{2}^-$, and $N(1535)\frac{1}{2}^-$ isobars have a significantly weaker influence on the results. Direct calculations show that the contribution of these resonances does not exceed $\sim 10\%$, which is in qualitative agreement with the conclusions of Ref.~\cite{SWA}. The meson-exchange currents and the $\Delta$-isobar excitation generate primarily dipole $E1$ and $M1$ transitions, and to a lesser extent, quadrupole $E2$ and $M2$ transitions. The role of higher multipoles ($E3$, $M3$, etc.) is even less significant and arises mainly from Fermi motion, which within the deuteron smears the dominant transitions into higher-order multipole components. Their influence becomes noticeable only in the immediate vicinity of the forward and backward angles, as one can see in Fig.~\ref{fig:T20_Mech}.

\section{Discussion of Results for $T_{20}$ in the Photon Energy Range $350$--$680$~MeV}
\label{sec:results}

\begin{figure}[t!]
\centering
\includegraphics[width=0.9\columnwidth]{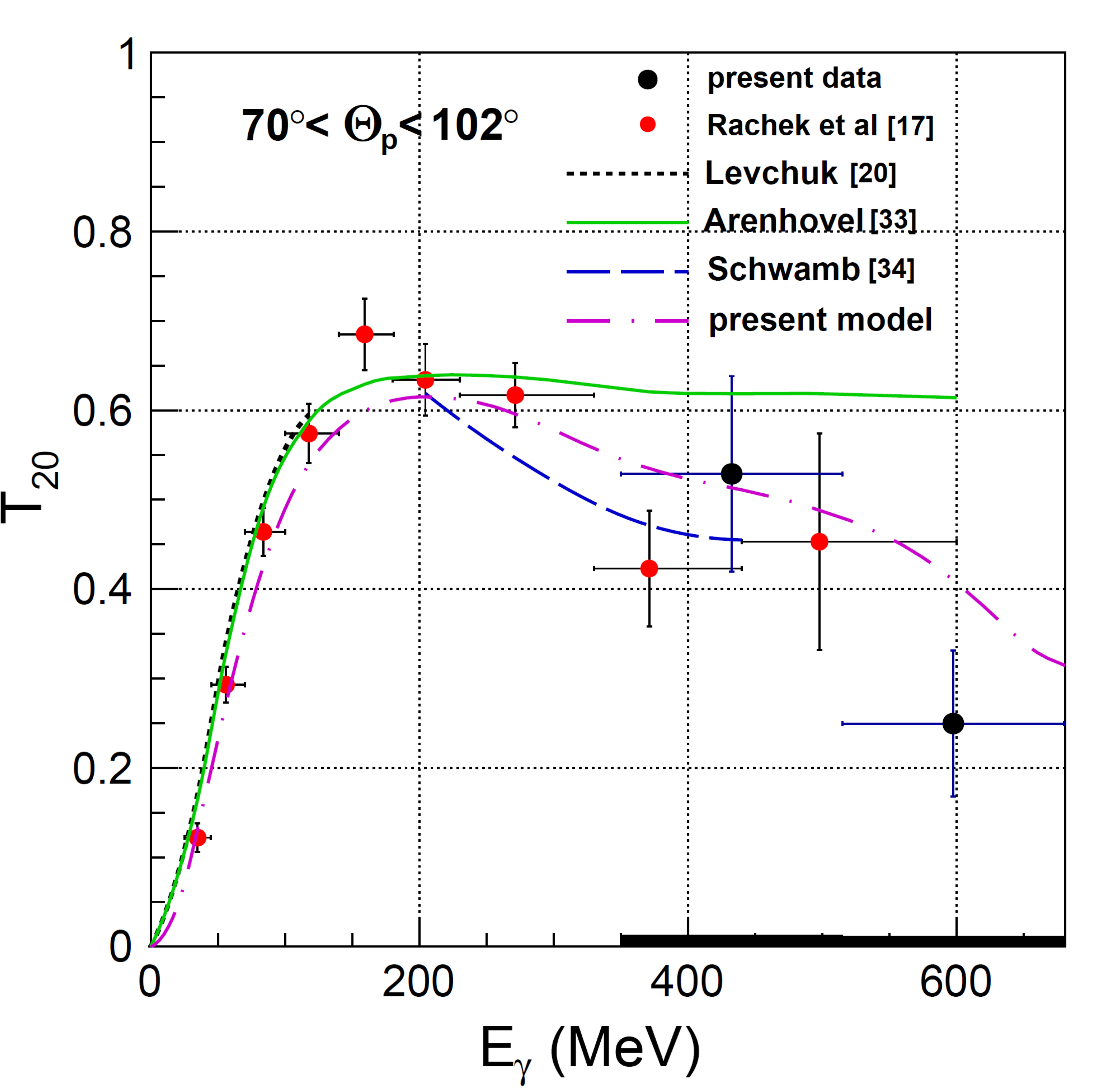}
\caption{
Tensor analyzing power $T_{20}$ in the $\gamma\vec{d}\to pn$ reaction as a function of the photon energy $E_\gamma$, averaged over the proton emission angle $\Theta_p = 70^\circ$--$102^\circ$. Black points: data from the present work; red points: results from Ref.~\cite{Rachek2007}. Only statistical uncertainties are shown. Curves: predictions of the models from Refs.~\cite{Levchuk} (black dotted), \cite{Schmitt} (green solid), and \cite{Schwamb} (blue dashed), and the model used in the present work (pink dash-dotted).
}
\label{fig:T20_data}
\end{figure}

Figure~\ref{fig:T20_data} shows the tensor analyzing power component $T_{20}$, averaged over the angular range $\Theta_p \in [70^\circ, 102^\circ]$. As can be seen, it remains positive across the entire energy range and demonstrates a gradual decrease from $\approx 0.65$ at $E_\gamma = 150$--$200$~MeV to $\approx 0.35$ in the $500$--$600$~MeV region.

The theoretical curves in Fig.~\ref{fig:T20_data} were obtained using approaches that include the one-body (nucleon) current based on the Bonn $NN$-potential~\cite{OBEP}, pion exchange currents (explicitly or via Siegert operators for electric multipoles), and electromagnetic transitions with the excitation of nucleon isobars, primarily $\Delta(1232)\frac{3}{2}^+$. The model~\cite{Schmitt} (solid green curve) additionally accounts for $\Delta\Delta$-components of the deuteron wave function and first-order relativistic corrections. The calculations~\cite{Schwamb} (dashed blue curve) were performed within the coupled-channels scheme ($NN$, $N\Delta$, $\pi d$), ensuring three-body $\pi NN$-unitarity in the isospin state $I=1$. Compared to the model~\cite{Schmitt} and the model used in this work, the approach~\cite{Schwamb} more consistently accounts for the opening of the $\pi NN$ channel above $E_\gamma \approx 140$~MeV, particularly the retardation effects and the gauge invariance of the full current.

As seen in the figure, in the energy region $E_\gamma \lesssim 150$~MeV, all the considered models yield similar results and are in good agreement with the experimental data~\cite{Rachek2007}, confirming that the low-energy dynamics are well understood. With increasing energy, the calculations become significantly more complex due to the opening of the three-body $\pi NN$ channel, which, in particular, leads to a noticeable spread in the theoretical predictions. At the same time, as shown in Fig.~\ref{fig:T20_data}, the model of Ref.~\cite{Schwamb}, as well as our model presented in Sec.~\ref{sec:model}, provides somewhat better agreement with the data. For a more stringent test of the theoretical models, additional measurements with higher statistics and an extension of the energy range above $1$~GeV would be highly valuable.

Measurements around $E_\gamma \approx 1$~GeV are also highly desirable, as they could help resolve certain fundamental difficulties in describing the $\gamma d \to np$ reaction above the resonance region. In particular, as shown in Ref.~\cite{wije01}, experimental data for the recoil nucleon polarization components $C_x$ and $C_z$ do not agree with the predictions of perturbative QCD (pQCD) at photon energies $1< E_\gamma < 2.4$~GeV. At the same time, meson-baryon models (in particular, the fully covariant model of Ref.~\cite{kang}) fail to reproduce the measured recoil polarization $P_y$ already at photon energies below 1~GeV. These findings imply that in the energy region $E_\gamma \sim 0.7\text{--}2.4$~GeV, deuteron photodisintegration could lie beyond the range of applicability of both purely quark-based (pQCD) and purely hadronic (meson-baryon) approaches. This in turn may reflect the transitional nature of this energy regime and suggest the need for hybrid theoretical frameworks.

In this context, it should be noted that the observable $P_y$ has traditionally been poorly described in the energy region above the first resonance. In particular, anomalously high values of $|P_y| \approx 0.6\text{--}0.8$ in the range $E_\gamma = 400\text{--}600$~MeV~\cite{Ikeda:1978ika} previously sparked discussions regarding the role of dibaryon resonances. The underlying cause of these discrepancies has not yet been definitively established. Therefore, the deviations in the behavior of $P_y$ at higher energies, considered in Ref.~\cite{wije01}, are unlikely to be conclusive evidence of the inadequacy of the meson-baryon picture. From this perspective, the satisfactory agreement of the tensor analyzing power $T_{20}$ with the theoretical predictions shown in Fig.~\ref{fig:T20_data} can be viewed as an additional argument supporting the adequacy of the meson-baryon approach (at least, within the current experimental uncertainties).

As predicted by perturbative QCD, the reaction $\gamma d \to pn$ should obey the hadron helicity conservation (HHC) rule in the limit $s \gg \Lambda_{\text{QCD}}^2$ at fixed $\Theta_p$~\cite{BrodskyHiller1992,Farrar1994}. The longitudinal polarization vector of the ultrarelativistic deuteron grows as $\sim \sqrt{s}/M_d$ in the c.m. frame, whereas the transverse ones remain of $\mathcal{O}(1)$. In the leading order of the $1/s$ expansion, where configurations with the minimum number of partons and $L_z = 0$ dominate, the hard amplitude does not compensate this enhancement. The helicity amplitudes with $\lambda_d = \pm 1$ are therefore power-suppressed relative to $\lambda_d = 0$, and asymptotically the reaction proceeds only through the longitudinal deuteron state. The cross section for the transverse target polarization ($P_{zz} = +1$) hence vanishes, and Eq.~(\ref{eq:dsigma_simple}) yields
\begin{equation}
\lim_{s \to \infty} T_{20} = -\sqrt{2}\,.
\label{eq:T20_pQCD}
\end{equation} 

Note that the overall trend exhibited by our experimental results (a monotonic decrease of $T_{20}$) is broadly consistent with the prediction (\ref{eq:T20_pQCD}). It is also evident that at our energies the behavior of $T_{20}$ is significantly influenced by effects not accounted for in the asymptotic regime (such as meson-exchange currents, finite quark masses, final-state interactions, etc.). In this context, it is instructive to analyze the sensitivity of the tensor analyzing power to the relative momentum spectrum of the nucleons in the deuteron within the energy region considered, $350 \leq E_\gamma \leq 680$~MeV. 

For a quantitative assessment, we employed the following method. In the calculations involving diagrams $(b)\text{--}(h)$ in Fig.~\ref{fig:diagrams}, the integration over the momentum $p$ at the $d\to pn$ vertex was restricted to the condition $p < p_c$ by introducing the step function $\theta(p_c - p)$ into the integrand, where $p = \frac{1}{2}\left|\vec{p}_p - \vec{p}_n\right|$ is the relative momentum of the nucleons in the deuteron. At this stage, the single-nucleon current, which does not contain loop integrals and contributes only marginally to the result, was neglected.

To quantify the sensitivity of $T_{20}$ to the internal deuteron momenta $p$, we used the normalized root-mean-square deviation $R(p_c)$, defined as
\begin{equation}
R(p_c)=\frac{\sqrt{\int\limits_{\Theta_1}^{\Theta_2}\left[d\sigma/d\Omega_p-d\sigma_c/d\Omega_p\right]^2\sin\Theta_p d\Theta_p}}
{\sqrt{\int\limits_{\Theta_1}^{\Theta_2}\left[d\sigma/d\Omega_p\right]^2\sin\Theta_p d\Theta_p}}\,,
\label{eq:R}
\end{equation}
where $d\sigma/d\Omega_p$ is the cross section from Eq.~(\ref{eq:dsigma_simple}), $d\sigma_c/d\Omega_p$ is the same cross section calculated with a momentum cutoff $p_c$, as described above, and the integration is performed over the measured angular range $\Theta_1=70^\circ$ to $\Theta_2=102^\circ$. The function $R(p_c)$ monotonically decreases from $1$ (at $p_c=0$) to $0$ (as $p_c \to \infty$), reflecting the contribution of momenta $p > p_c$ to the formation of $T_{20}$.

To determine the characteristic momentum scale and the boundaries of the sensitivity interval, we analyze the integrated sensitivity of the function $R(p_c)$. For this purpose, we introduce a normalized rate of change
\begin{equation}
w(p_c) = \frac{1}{V_{\mathrm{total}}} \left| \frac{dR}{dp_c} \right|, \quad V_{\mathrm{total}} = \int_0^\infty \left| \frac{dR}{dp_c'} \right| dp_c',
\end{equation}
which serves as a weight function representing the relative contribution of each momentum $p_c$ to the total variation of the observable. The characteristic momentum $p_{\mathrm{char}}$ is defined as the point where the accumulated change $\int_0^{p_{\mathrm{char}}} w(p_c')\,dp_c'$ reaches 50\% of the total. The boundaries of the sensitivity interval $[p_{<}, p_{>}]$ correspond to the 5\% and 95\% levels of this accumulated change. This choice of thresholds is analogous to the standard truncation of tails in spectral analysis. The resulting interval $[p_{<}, p_{>}]$ characterizes the range of internal deuteron momenta to which the observable $T_{20}$ is most sensitive.

\begin{figure}[t!]
\centering
\includegraphics[width=1.0\columnwidth]{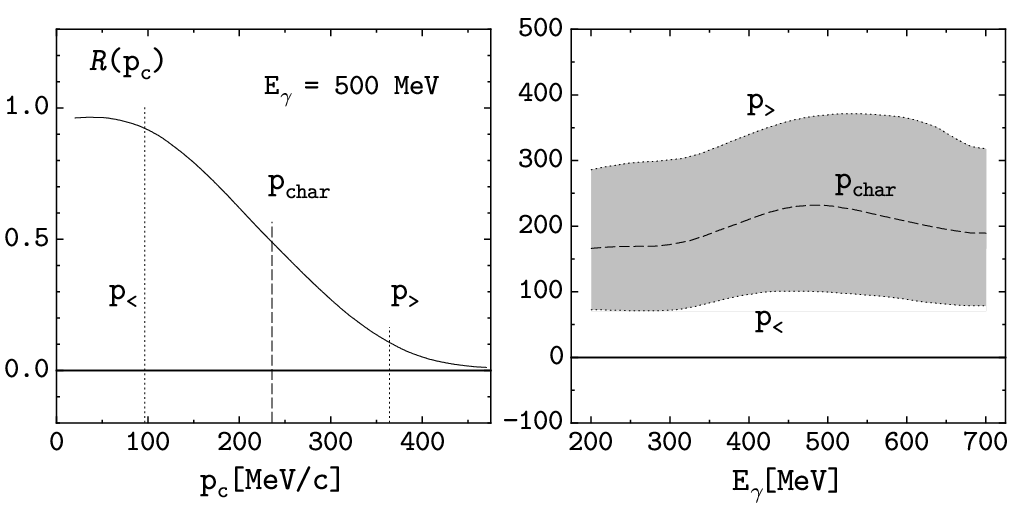}
\caption{
Left panel: Sensitivity function $R(p_c)$ [Eq.~(\ref{eq:R})] for the tensor analyzing power $T_{20}$ at $E_\gamma = 500$~MeV. The vertical dashed lines denote the characteristic momentum $p_{\rm char}$ (where 50\% of the total variation is accumulated) and the boundaries of the sensitivity interval $p_<$ and $p_>$ (corresponding to 5\% and 95\% of the cumulative variation). Right panel: Energy dependence of the characteristic momenta. The dashed curve shows $p_{\rm char}(E_\gamma)$, and the dotted curves represent the boundaries of the sensitivity interval $p_<(E_\gamma)$ and $p_>(E_\gamma)$. The shaded region indicates the range of internal momenta to which the $T_{20}$ measurement is most sensitive at a given photon energy.
}
\label{fig:R_vs_pc}
\end{figure}

The characteristic behavior of $R(p_c)$ [Eq.~(\ref{eq:R})], calculated with our model (Sect.~\ref{sec:model}), is shown in Fig.~\ref{fig:R_vs_pc} (left panel). The right panel displays the energy dependence of the characteristic momentum $p_{\rm char}$ and the boundaries of the sensitivity interval $[p_<, p_>]$ as a function of the photon energy $E_\gamma$. Notably, the boundary values $p_<$ and $p_>$ vary only weakly with increasing energy. Across almost the entire energy range, the dominant contribution to the $T_{20}$ asymmetry arises from relative momenta in the range $[70, 300]$ MeV/$c$, which, by the uncertainty principle, corresponds to internucleon distances of approximately 0.7 to 3 fm. Clearly, under these kinematic conditions, the overlap of the individual nucleon wave functions is not very important, which should support the applicability of the meson-baryon approach.

The results presented here also indicate that, even in the region above the first resonance, the $\gamma d \to np$ reaction is predominantly sensitive to the low-momentum components of the deuteron wave function. First, this provides \textit{a posteriori} justification for our use of the separable representation of the Paris potential~\cite{Haiden}, whose range of applicability, as noted earlier, is limited to $NN$ laboratory energies up to 300 MeV. In other words, although contributions from higher momenta are included in the calculations, their impact on the final results is limited. Second, our $T_{20}$ measurements probe the spatial structure of the deuteron at the same characteristic distances as the unpolarized differential cross section. This constitutes a fundamental difference between our measurements and experiments on deuteron electrodisintegration ($ed \to e'pn$) at $Q^2 > 0$, where an increase in the three-momentum transfer allows one to sequentially probe the deuteron wave function at increasingly higher relative momenta. In contrast, our results should be viewed primarily as a source of information on the mechanisms of deuteron photodisintegration, particularly the dynamics of meson-exchange currents and the excitation of nucleon resonances.

\section{Conclusion}
\label{sec:conclusion}

The measurement of the tensor analyzing power $T_{20}$ in the reaction $\gamma \vec{d} \to pn$ at photon energies $E_\gamma = 350$--$680$~MeV was carried out as part of a precision measurement program at the VEPP-3 storage ring. The program is aimed at improving the accuracy of experimental data and rigorously testing theoretical models above the pion production threshold. To achieve this, a photon tagging system and a wide-aperture sandwich detector for neutrons were implemented in the experimental setup. The obtained results are generally consistent with previously published data~\cite{Rachek2007} at $E_\gamma < 600$~MeV.

The measured tensor analyzing power $T_{20}$ is positive and is qualitatively described by the meson-baryon model, which is dominated by meson-exchange currents and $\Delta$-isobar excitations. However, the current statistical uncertainty does not yet allow for an unambiguous assessment of the quantitative agreement with theoretical predictions.

As repeatedly emphasized in the preceding sections, some currently available experimental data suggest the possible breakdown of the conventional meson-baryon picture already at photon energies of $\sim 1$~GeV. However, since we are still far from the perturbative QCD regime~\cite{BrodskyHiller1992}, before drawing conclusions about the role of quark-gluon degrees of freedom, it is necessary to minimize the theoretical uncertainties of the meson-baryon models themselves. Crucial tasks include a rigorous covariant treatment of relativistic effects (including approaches based on the Bethe-Salpeter equation and/or the covariant model developed by the Bonn group~\cite{kang}), as well as the consistent implementation of three-body unitarity with a correct description of true pion absorption mechanisms~\cite{Rinat:1978sa}. Only after establishing a consistent framework for describing the reaction that accounts for all major corrections can residual discrepancies be unambiguously attributed to subnucleon structure effects, thereby justifying the transition to hybrid models that combine traditional nuclear degrees of freedom with parton dynamics~\cite{Kisslinger:1986iw}. It is hoped that the $T_{20}$ data will expand the experimental database for testing these approaches and, together with other measurements, help to pinpoint the kinematic region where the traditional meson-baryon description begins to exhibit systematic deviations.

The future development of the experimental program at VEPP-3 involves a substantial increase in accumulated statistics and an extension of the photon energy range above 1~GeV. This will help clarify the discrepancies between theory and experiment, as well as trace their evolution with increasing energy. For a comprehensive test of the models and a more detailed understanding of the reaction dynamics, we also intend to measure the $T_{21}$ and $T_{22}$ components, which are sensitive to different interference combinations of the transition amplitudes. The expected results will not only allow for the refinement of existing models but may also quantitatively determine the limits of applicability of the traditional meson-baryon picture at large momentum transfers, thereby laying the groundwork for precision tests at next-generation facilities.


\begin{acknowledgments}
Financial support for this work was provided by the Ministry of Science and Higher Education of Russian Federation, Science Program No. FSWW-2023-0003.
\end{acknowledgments}


\end{document}